\documentclass[referee]{raa}            

\usepackage{graphicx,times}             
\usepackage{natbib}
\usepackage{amssymb,amsmath}
\bibpunct{(}{)}{;}{a}{}{,}

\usepackage[pagebackref=true]{hyperref}
\usepackage{amsmath}
\usepackage{multirow}
\usepackage[ruled]{algorithm2e}
\usepackage{diagbox}
\usepackage{threeparttable}
\usepackage{gensymb}
\usepackage{indentfirst}

\begin{document}

  \title{A Modern ConvNet for Solar Filament Detection
}


   \author{J.R. Hu 
      \inst{1}
   \and Q. Hao
      \inst{1,2,*}\footnotetext{$*$Corresponding Author}
   \and Z. Zheng
    \inst{1,2}
   \and P. F. Chen
    \inst{1,2}
   \and C. Li
    \inst{1,2,3}
   \and Y. Meng
    \inst{1}
   }

   \institute{School of Astronomy and Space Science, Nanjing University, Nanjing 210023, China; {\it haoqi@nju.edu.cn}\\
        \and
             Key Laboratory of Modern Astronomy and Astrophysics, Ministry of Education, Nanjing 210023, China\\
        \and
             Institute of Science and Technology for Deep Space Exploration, Suzhou Campus, Nanjing University, Suzhou 215163, China\\
\vs\no
   {\small Received 202x month day; accepted 202x month day}}

\abstract{Automated solar filament detection using deep learning faces several challenges. Semantic segmentation of solar filaments is a complicated multiscale feature extraction task with long-tail distribution. Furthermore, a large-scale, highly complete, and finely detailed dataset has become mandatory for providing abundant information. To address these challenges, we present a series of machine learning approaches to develop a solar filament detection workflow that performs superbly. First, we manually annotated a small-scale solar filament dataset based on H$\alpha$ spectra called MHAS. Next, we developed the Multiscale ORiented DENdritic (MORDEN) model, a semantic segmentation model focusing on multiscale feature extraction. We also introduced the Dense Conditional Random Field (DenseCRF) and Density-Based Spatial Clustering of Applications with Noise (DBSCAN) methods for post-processing. Using the proposed workflow, we generated a large-scale, high-quality dataset called AHAS. Experimental results demonstrate that MORDEN outperforms several existing solar filament semantic segmentation models with open access. DenseCRF has been demonstrated to effectively capture fine edge details. We also evaluated the effects of data scaling and the reliability of DBSCAN and found that both approaches yield satisfactory performance. Multiple visualization results substantiate our quantitative findings. Our work provides a foundation for maximizing the potential of deep learning models for solar filament detection.
}

   \authorrunning{J.R. Hu et al.}            
   \titlerunning{A Modern ConvNet for Solar Filament Detection}  

   \maketitle

%
%

\section{Introduction}
\label{sec:1}

Solar filaments, also called prominences when observed above the solar limb, are intriguing magnetized structures composed of cool, dense plasma suspended within the hot tenuous corona \citep{Tandberg1995}. Typically, filaments are 100 times cooler and denser than surrounding corona \citep{parenti14, Vial2015}. They are particularly visible in H$\alpha$ observations on the solar disk, typically manifesting as elongated dark structures with multiple barbs \citep{Martin1998}. Filaments usually appear along the magnetic polarity inversion line and extend in the solar atmosphere from the chromosphere to the corona \citep{Vial2015}, and these features allow us to track and analyze the solar magnetic field by studying the long-term evolution of a large number of filaments \citep{Li2010,Hao2015,Gopalswamy2016,Zhang2024}. In addition, filaments occasionally undergo large-scale instabilities that break their equilibrium and result in eruptions. This suggests a correlation with solar flares and coronal mass ejections (CMEs), which are considered the key drivers of hazardous space weather. \citep{Chen11,Gopalswamy2015,chen20}. Therefore, it is of great importance and significance to carry out statistical analyses of filaments as well as case studies.

In recent years, there have been rapid advancements of observational capabilities, resulting in a significant increase in data quality and quantity. A huge volume of high cadence and resolution data has been collected, posing a challenge for efficient processing. To overcome the limitations of human resources and time cost, automated detection has taken the mainstream to derive the features of interest in the observations for further solar filament analysis. Additionally, the quality of observations from different observatories over the past decades varies, which also emphasizes the importance of robust analytic procedures. Achieving ideal performance for future observations while reliably adapting to historical data remains a major challenge.

Filament detection methods applying classical image processing focus on the intensity of solar filaments. In full-disk H$\alpha$ images, filaments are characterized as dark structures with high contrast against their surroundings. Consequently, the typical way is to extract filaments by setting an appropriate threshold, either a global threshold \citep{Gao2002} or a local threshold \citep{Shih2003,Fuller2005,Bernasconi2005} combined with the region-growing method. However, it is impractical to rely on a universal threshold because the intensity of filaments varies significantly due to several factors such as exposure time and seeing conditions. In order to solve this problem, intensity normalization and adaptive threshold methods were adopted \citep{Qu2005,Yuan2011,Hao2015}. Still, these methods lack robustness sometimes due to complex scenarios associated with the variations of instruments and observing conditions.

Deep learning methods are given preference in recent years, and three types of workflow have been developed. Solar filament detection can be decomposed into two subtasks: morphological extraction and fragment integration. Which subtask to begin with determines the type of the workflow. (1) Some researchers concentrated on the morphology characteristics of solar filaments \citep{Zhu2019,Liu2021,Zheng2024,Jiang2024,Zhu2025}, and therefore proposed semantic segmentation models based on U-Net \citep{Ronneberger2015}, a kind of hierarchical convolutional neural networks (CNNs). Semantic segmentation models are well designed for morphological recognition, but will not distinguish different filaments. As a follow-up, a clustering algorithm is also utilized as post-processing for fragment integration \citep{Zheng2024}. (2) \citet{Diercke2024} proposed another two-step workflow which identifies each filament first. They applied YOLOv5 \citep{Jocher2020} which detect the bounding boxes of filaments, followed by a U-Net for downstream semantic segmentation. (3) In contrast, \citet{Guo2022} finished the two subtasks with only one instance segmentation model CondInst \citep{Tian2020,Tian2023} that directly allocates identifier for every filament at the pixel level. These three workflows are all demonstrated to be effective. Nevertheless, fragment integration based on solitary H$\alpha$ images is regarded as semantically ambiguous, which relies exclusively on morphological characteristics, such as the distance, slope, and similarity among the fragments \citep{Hao2018}. This indicates that neither humans nor deep learning models can accurately identify which fragments belong to a single filament without referring to other bands, such as photospheric magnetic polarity inversion lines (PILs) in magnetograms or filament channels in 304 \AA \ filtergrams, where filaments are always aligned \citep{Tandberg1995,Martin1998,Chen2025}. We are concerned that the annotated data for fragment integration may introduce individual bias and confuse the deep learning models, consequently causing influences on morphological extraction difficult to quantify. Therefore, the first type of workflow, which decouples the two subtasks to ensure reliable morphological information, is preferred from our perspective.

Additionally, solar filament semantic segmentation is a typical multiscale feature extraction task with a long-tail distribution. Filament size can vary greatly. Active region filaments are relatively small with the length less than 10$^4$ km, while the quiescent filament can be relatively large with typical length around 10$^5$ km \citep{Vial2015,Hao2015,Zhang2024}. Furthermore, quiescent filaments exhibit more complex morphological features on small scales, such as filament barbs \citep{Tandberg1995}. Filaments also commonly occupy less than ten percent of the solar disk, resulting in an uneven distribution between the target and the background, known as a long-tail distribution \citep{Hao2015}. Many innovations have been adopted to address these two problems. \citet{Liu2021} enhanced the receptive field of U-Net by applying Atrous Spatial Pyramid Pooling (ASPP, \citealt{Deeplab2018}) and appending extra skip connections. \citet{Jiang2024} and \citet{Zhu2025} mainly introduced Channel Attention Mechanism (CAM) and Spatial Attention Mechanism (SAM) \citep{CBAM2018} for multiscale feature extraction. \citet{Zheng2024} mitigated the long-tail distribution with focal loss \citep{Lin2017}. While these innovations remain valid, they have been somewhat overshadowed by the recent rapid development in computer vision. As a result, we propose Multiscale ORiented DENdritic model (MORDEN), which combines the advantages of these methods and makes relevant ``modern'' refinements to achieve ideal segmentation performance in complex observational environments. Specifically, we enlarge the kernel size of the convolutional layers and replace the Batch Normalization (BN, \citealt{BatchNorm2015}) in the downsampling stage with Instance Normalization (IN, \citealt{InstanceNorm2017}). We also develop several multiscale oriented modules including Adaptive Pyramid Pooling (APP), Patch Attention Gate (PAG), branch block, and multihead structure.

High-quality datasets are as essential as well-designed models. A well-annotated dataset provides deep learning models with abundant valid semantic information \citep{imagenet}. Existing methods have suggested some approaches for data annotation. For example, manual selection and automated correction of results from traditional methods facilitate the creation of large-scale datasets \citep{Zhu2025,Jiang2024}. Manual selection incorporates a certain amount of prior knowledge, enabling more precise semantic information in the datasets. However, some tiny filaments not extracted by traditional methods are still excluded from the training set. Deep learning models will fail to extract these finer filaments since they are not informed during training. Conversely, although manual annotation ensures that all visible filaments are included, the shortage of professional manpower makes it difficult to obtain large-scale datasets. To address this dilemma, we expect the filament detection method to do more than just distill a dataset annotated by humans or computers. We need a model that performs well under the current conditions and a workflow that further allows us to construct large-scale, highly complete, and finely detailed datasets for future applications. This is crucial for deploying Transformer-based models \citep{ViT2021,SwinT2021}, as large-scale datasets with abundant information will be a mandatory requirement for these models. Consequently, we employ Dense Conditional Random Field (DenseCRF, \citealt{DenseCRF2011}) for post-processing, which optimizes edge details with a high degree of credibility.

In this paper, we propose a filament detection workflow that consists of the semantic segmentation model MORDEN and two post-processing approaches, DenseCRF and Density-Based Spatial Clustering of Applications with Noise (DBSCAN, \citealt{Schubert2017}). MORDEN is designed for multiscale feature extraction with several structural innovations. DenseCRF optimizes edge details, and DBSCAN integrates filament fragments. Section~\ref{sec:2} introduces the comprehensive methodology of the workflow. Section~\ref{sec:3}  demonstrates the results of experiments designed to validate the workflow. In Section~\ref{sec:4}, we discuss the phenomena in each experiment and their underlying mechanisms. Finally, a succinct conclusion is presented in Section~\ref{sec:5}.

\begin{figure*}
    \centering
    \includegraphics[width=1.0\linewidth]{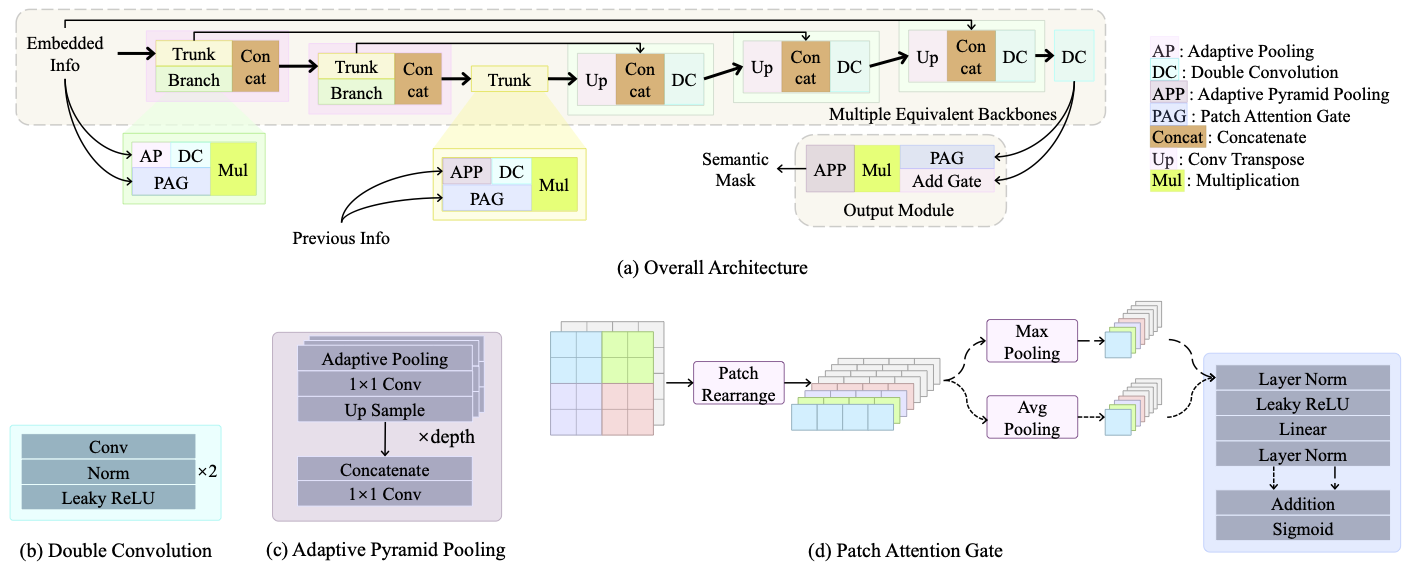}
    \caption{Overall architecture and primary modules of MORDEN.}
    \label{fig:1}
\end{figure*}

\section{Method}
\label{sec:2}
\subsection{MORDEN: Multiscale Oriented Dendritic Semantic Segmentation Model}
\label{sec:2.1}
\subsubsection{Modern Refinements}
\label{sec:2.1.1}

The kernel size is considered a critical factor in the receptive field of CNNs. However, expanding the kernel size also increases training difficulty, parameter counts, and computational demands \citep{ConvNext2022}. Considering computational resources, we increase the kernel size of U-Net from 3 to 5. Furthermore, Instance Normalization (IN) is considered to be more sensitive to texture features and relative color information than Batch Normalization (BN) \citep{AdaIN2017}. In other words, IN focuses more on local contrast variations, while BN retains the global intensity statistics. This property is particularly suitable for solar filament segmentation because local intensity variations are critical, and even decisive, indicators on H$\alpha$ images.  We also apply some common optimizations, such as leaky ReLU \citep{maas2013rectifier} and Dropout \citep{Dropout2014}. Specifically, a dropout layer follows every module in Figure~\ref{fig:1}.

\subsubsection{Adaptive Pyramid Pooling}
\label{sec:2.1.2}

Most of the pyramid-shaped modules designed for multiscale feature extraction, such as ASPP and Pyramid Pooling Module (PPM, \citealt{Zhao2017}), are intended for high-dimensional representations in neural networks. However, they cannot be applied to every block in a hierarchical model due to their fixed target pooling size. To mitigate information loss in each downsampling block, we modify PPM into Adaptive Pyramid Pooling (APP) to accommodate feature maps of any size. Figure~\ref{fig:1}(c) shows the modified pyramid pooling structure, where the skip connection in PPM is removed to allow adaption to different output sizes. Furthermore, the output sizes of the adaptive pooling layers are no longer static, but are determined by the final output size of APP module as follows:
\begin{equation}
    \begin{aligned}
        \mathit{S} = \{ (h_i, w_i) \; | \; 
        & h_i = \max ( \lfloor \frac{\mathit{H} \cdot i}{\mathit{L}} \rfloor , 1 ) , \\
        & w_i = \max ( \lfloor \frac{\mathit{W} \cdot i}{\mathit{L}} \rfloor , 1 ) ,
        \;i = 1, 2, \dots, \mathit{L}
        \},
    \end{aligned}
\end{equation}
where $\mathit{H}$ and $\mathit{W}$ denote the height and width of the output size respectively, $\mathit{L}$ stands for the number of adaptive pooling layers in APP module, and $h_i$ and $w_i$ are the target height and width of each pooling layer.

APP inherits the advantages of PPM in terms of few training parameters, and the arithmetic progression format broadens the applicability of PPM. Unlike PPM and ASPP, which are only useful for integrating high-level information, APP can be used at any stage of a hierarchical model. This allows for better representation of low-level information and benefits the information pathway in the skip connections of U-Net based models. It functions by scaling diverse pooling sizes for multiscale feature fusion when dealing with large low-level feature maps, and it shares similar structure with PPM when processing small high-level representations. Notably, the skip connection in PPM is a special case equivalent to APP when the output size equals to the input size. As shown in Figure~\ref{fig:1}(a), APP replaces the plain pooling layer of the light yellow "trunk" block which comes from the original downsampling block, and also contributes to the output module.

\subsubsection{Patch Attention Gate}
\label{sec:2.1.3}

Channel attention mechanism (CAM) controls the information interaction between different feature channels. Spatial attention mechanism (SAM) highlights the most relevant locations by generating spatial weights. However, neither of them possesses the capability of global attention, mainly constrained by the finite receptive field of the global pooling and the convolutional layer. Conversely, the rearrangement of image patches in Vision Transformer (ViT, \citealt{ViT2021}) has been verified to be an effective method for global attention. This inspired the development of our proposed Patch Attention Gate (PAG). 

As is shown in Figure~\ref{fig:1}(d), the feature map is split into multiple small patches. Then, each patch is compressed into one value using two pooling strategies. This facilitates the generation of channel weights with a linear layer. This structure recognizes both channel features and global spatial information at a relatively finer scale. Additionally, to optimize the gating mechanism, PAG  supplements extra information flow as illustrated in Figure~\ref{fig:1}(a). In other words, PAG can be formulated as $\mathcal{G}(x_1, x_2) = g(x_1) * x_2$, where the low-level context $x_1$ serves as an additional information source, and provides general regulation of high-level feature maps $x_2$. To balance information abundance and computational complexity, given the input size $S_{x_1}$ for function $g$ characterized by height and width within powers of 2, the patch size is specified by the following formula:
\begin{equation}
    S_{patch} = 2^{\;\lfloor  \log_2\sqrt{S_{x_1}}  \rfloor}.
\end{equation}

\subsubsection{Branch Block}
\label{sec:2.1.4}

Hierarchical skip connections are highly effective at reintroducing pixel-wise semantic detail from downsampling stage to upsampling stage. This reflects the information bottleneck in plain CNNs. Inspired by this, we introduce hierarchical skip connections from the embedded information to the downsampling stage, named branch blocks. Unlike sequential trunk blocks which contribute to the primary information flow, branch blocks are shortcuts that embed information into different scales, reminding the model of lost information during progressive encoding. The light green frame illustrated in Figure~\ref{fig:1}(a) shows the composition of one branch block, which consists of an adaptive pooling layer, a double convolution module and a PAG. Through these layers, structural semantic features are transformed into arbitrary sizes.

\subsubsection{Multihead Structure and Output Module}
\label{sec:2.1.5}

Multihead structure is widely recognized for its high training efficiency and strong robustness \citep{Vaswani2017}. It divides one task into multiple subspaces, each of which shares the equivalent potential to achieve optimal convergence as a standalone model. These heads are also theoretically fully parallelizable. Meanwhile, the gating mechanism is considered significant for filtering redundant context and broadening the information pathway \citep{QwenGate2025}. As Figure~\ref{fig:1}(a) shows, MORDEN consists of multiple identical U-shaped backbones and a unified output module. The output module employs additive gating and multiplicative gating to consolidate semantic representations from different backbones. Then, an APP integrates multiscale features and produces the semantic mask. The additive gating is equivalent to a residual connection with a double convolution module, while the multiplicative gating uses PAG.

\subsection{DenseCRF: Post-processing for Finer Detail}
\label{sec:2.2}

DenseCRF is well adapted for post-processing in semantic segmentation tasks \citep{DenseCRF2011,Deeplab2018}. It constructs a conditional random field $(\mathbf{I}, \mathbf{X})$ where $\mathbf{I}$ ranges over possible input images and $\mathbf{X}$ ranges over possible labelings at pixel-level. \citet{DenseCRF2011} simplified the conditional probability model $P(\mathbf{X} | \mathbf{I})$ with Gibbs distribution as follows:

\begin{equation}
    P(\mathbf{X} | \mathbf{I}) =
    \frac{1}{Z(\mathbf{I})} \;
    \mathrm{exp} \; \bigl(-E(\mathbf{x} | \mathbf{I})\bigr),
\end{equation}

where the Gibbs energy $E(\mathbf{x})$ is defined with a unary potential and a pairwise potentials:

\begin{equation}
    E(\mathbf{x}) =
    \sum_{i} \psi_u(x_i) +
    \sum_{i < j} \psi_p(x_i, x_j).
\end{equation}

$\psi_u(x_i)$ denotes the distribution over the label assignment $x_i$ given corresponding pixel features, which can be provided by the semantic segmentation model. $\psi_p(x_i, x_j)$ represents the relationship of $x_i$ and $x_j$ under their pixel feature correlations, mainly decided by their positions and intensities:

\begin{equation}
        \psi_p(x_i, x_j) = 
        \mu(x_i, x_j)\;k(\mathbf{f}_i, \mathbf{f}_j),
\end{equation}
\begin{equation}
    \mu(x_i, x_j) = [x_i \neq x_j],
\end{equation}
\begin{equation}
    \begin{aligned}
        k(\mathbf{f}_i, \mathbf{f}_j) =\;
        & \omega^{(1)} \mathrm{exp} \Bigl(
        -\frac{|p_i-p_j|^2}{2\theta^2_\alpha}
        -\frac{|I_i-I_j|^2}{2\theta^2_\beta}\Bigr) \;+ 
        & \omega^{(2)} \mathrm{exp} \Bigl(
        -\frac{|p_i-p_j|^2}{2\theta^{2}_\gamma}
        \Bigr),
    \end{aligned}
    \label{pairwise-potentials}
\end{equation}
where $\mu(x_i, x_j)$ is a label compatibility function that introduces a penalty on nearby similar pixels assigned with different labels, and $p$ and $I$ are the position and intensity of each pixel respectively. In other words, the likelihood of arbitrary two pixels being assigned the same label increases with their spatial proximity and intensity similarity. This matches the characteristics of solar filament perfectly and meets the need for refining edge details. While MORDEN is reinforced for structural multiscale feature extraction at a global scale, DenseCRF further focuses on the local intensity features of each identified filament fragment.

Beyond finer detail, DenseCRF resolves the limitation of CNNs that one model cannot perform superbly under arbitrary spatial resolution. Additionally, the cost of training resources and model size will increase with the pixel resolution of the image. Due to the pairwise potentials based on the original image, high resolution intensity information can also be incorporated into the post-processing stage. Furthermore, DenseCRF mitigates the "black box model" problem and data distillation. Since MORDEN only contributes to the unary potential, introducing original information and the probabilistic modeling which are independent of the model's predictions also plays a crucial role in the final reasoning process. This indicates that the final output is not merely an approximation of the manually annotated training data but rather a posterior probability derived from multiple reliable sources, especially the intensity and spatial information in the original image. This paves the way for constructing large-scale and finely detailed datasets because integrating model-based location and image-driven detail refinement addresses the risk of self-referential annotation to some extent.

In addition to DenseCRF, we also perform hole filling with the Open Source Computer Vision Library (OpenCV, \citealt{opencv}) to ensure that every extracted fragment is a simply connected domain.

\subsection{DBSCAN: Fragment Integration}
\label{sec:2.3}

We employ the density-based spatial clustering of applications with noise (DBSCAN, \citealt{Schubert2017}) for fragments integration. In addition to distinguishing every solar filament, this clustering algorithm reduces the noise on the semantic mask caused by AI hallucinations, resulting in more robust and reliable detection.

\section{Result and Analysis}
\label{sec:3}
\subsection{Training Settings}
\label{sec:3.1}
\subsubsection{Data Preparation and Expansion}
\label{sec:3.1.1}

To evaluate the performance of our semantic segmentation model, we collected several open-access datasets from recent studies that primarily focused on semantic segmentation, and created a manually annotated dataset. Limited by the accessibility and differences in metrics used, the two-step detection workflow \citep{Diercke2024} and the instance segmentation approach \citep{Guo2022} were not included. We used our model, trained on the manually annotated dataset along with DenseCRF and hole filling, to generate a large-scale dataset. Detailed information about each dataset and its organization is provided below.

\textbf{ \citet{Jiang2024}}: The paper claims that this dataset contains 600 solar images, 100 of which are publicly available. First, the filaments were labeled using an automated clustering algorithm, k-means \citep{MacQueen1967}. Then, automated methods, including particle erosion, multi-closing operation, and hole filling, corrected the labels. Since the distribution of these 100 images is not specified, they were simply split into three sets: a training set of 90 images, a validation set of five images, and a test set of five images, ordered by observation time.

\textbf{ \citet{Zhu2025}}: This dataset primarily consists of a training set of 1,007 images and a test set of 98 images. The filaments were manually selected from annotated images using traditional methods. Due to the lack of a validation set, the last 100 images sorted by observation time from the training set were selected as the validation set.

\textbf{H$\alpha$ spectra based dataset (HAS, \citealt{Zheng2024})}: This dataset contains 120 labeled images manually selected from the k-means results. The images are divided into the training, the validation, and the test sets in a ratio of 4:1:1.

\textbf{Manual HAS dataset (MHAS)}: This dataset consists of 120 manually annotated images with the same ratio as HAS dataset. Based on the initial k-mean annotations, manual corrections were made to eliminate incorrect labels and provide more specific semantic information. Specifically, some dark filaments located above active regions were supplemented and the fragmented filaments within the same filament channel were manually connected. Additionally, the subtle boundary structures of low-contrast filaments that appear faint and are difficult to reliably identify through spectral-based classification alone, were also refined through manual annotation.

\textbf{Automated HAS dataset (AHAS)}: After training MORDEN on MHAS, we produced an automatically annotated large-scale dataset using our workflow. Specifically, we applied DenseCRF and hole filling on the output of the model, resizing it to the size of the original image size to integrate more original intensity and spatial information. We gathered the observations from Chinese H$\alpha$ Solar Explorer (CHASE; \citealt{Li2019,Li2022}), ranging from 2024 to 2025, at a cadence of one image per day. After manually screening the processed results, 695 annotations were retained. In addition, the 120 images from MHAS dataset were re-annotated using our workflow. This resulted in a final dataset totaling 775 images in the training set, 20 images in the validation set and 20 images in the test set. To be specific, the original images in the validation and the test set are identical to those in MHAS dataset.

We trained our model on a single RTX H100 and constructed a data augmentation pipeline that includes several transformations aligned with real-world scenarios. The corresponding augmentations and their probabilities and ranges are described in Table~\ref{table:1}

\renewcommand{\arraystretch}{1.0}
\begin{table}[ht!]   
\caption{Data augmentation pipeline.\label{table:1}}
\begin{center}
\begin{tabular}{c|c|c|c}    
\hline   
\textbf{Augmentation} & \textbf{Probability} & \textbf{Variable} & \textbf{Range} \\   
\hline   vertical flipping & 50\% &  & \\ 
\hline   rotation & 90\% & angle & [-45\degree, 45\degree] \\ 
\hline   gamma & \multirow{2}{*}{90\%} & \multirow{2}{*}{gamma} & \multirow{2}{*}{[0.5, 1.5]} \\
         transformation & & & \\
\hline   scaling & 50\% & ratio & [0.5, 1.0) \\
\hline   min-max & \multirow{2}{*}{100\%} & \multirow{2}{*}{} & \multirow{2}{*}{} \\
         normalization & & & \\
\hline   
\end{tabular} 
\end{center}   
\end{table}

\subsubsection{Hyperparameters}
\label{sec:3.1.2}

The datasets differ with each other in scale and image resolution, incurring different sets of hyperparamters for achieving optimal performance and metrics. Table~\ref{table:2} shows the hyperparameters of MORDEN, where base channels and mid channels refer to the number of embedding channels and output channels of each head, respectively. Branch channels refer to the output channels of each branch block. Leaky ReLU introduces an additional  hyperparameter which is the negative slope. Additionally, we used focal loss to address the long-tail distribution, which introduces two hyperparameters: $\alpha$ and $\gamma$. AdamW \citep{adamw} was utilized as the optimizer.


\renewcommand{\arraystretch}{1.0}
\begin{table*}[ht!]   
\caption{Hyperparameters for each dataset. \label{table:2}}  
\begin{center}
\begin{threeparttable}
\begin{tabular}{c|c|c|c}    
\hline    \textbf{\diagbox[width=17em]{Parameter}{Value}{Dataset}} & \textbf{xHAS\tnote{*}} & \textbf{\citeauthor{Zhu2025}} & \textbf{\citeauthor{Jiang2024}} \\
\hline base channels & 40 & 40 & 8 \\
\hline mid channels & 40 & 40 & 8 \\
\hline branch channels & 20 & 20 & 8 \\
\hline pooling depth & 5 & 4 & 5 \\
\hline number of heads & 2 & 2 & 3 \\
\hline dropout & 0.1 & 0.1 & 0.1 \\
\hline negative slope & 0.01 & 0.01 & 0.01 \\
\hline alpha & 0.66 & 0.66 & 0.66 \\
\hline gamma & 2 & 2 & 2 \\
\hline batch size & 3 & 3 & 2 \\
\hline learning rate & 2e-4 & 2e-4 & 2e-4 \\
\hline image height & 1024 & 512 & 512 \\
\hline image width & 1024 & 512 & 512 \\
\hline   
\end{tabular}

\begin{tablenotes}
    \footnotesize
    \item[*] xHAS denotes the three datasets with names ending with HAS.
\end{tablenotes}
\end{threeparttable}
\end{center}   
\end{table*}

DenseCRF depends mainly on the parameters described in Formula~\ref{pairwise-potentials} and on iterations, which are adjusted to fit different image sizes, as shown in Table~\ref{table:3}.

DBSCAN also introduces two hyperparameters, $\epsilon$ and minPts. $\epsilon$ indicates the distance threshold between different clusters, and minPts indicates the minimum number of points that each cluster should contain. Through manual judgment on the validation set, we set $\epsilon$ to 45, and minPts to 140. Considering that the typical spatial resolution of CHASE data is 1.04$\arcsec$ per pixel, the upper bound of the distance between filament fragments is approximately 33,930 km, and the lower bound of the area of a filament should be about $8\times10^{7}\ \mathrm{km^{2}}$. For actual deployment, these two hyperparameters should be adjusted according to the spatial resolution of the input data. 

\renewcommand{\arraystretch}{1.0}
\begin{table}[ht!]
\caption{Hyperparameters of DenseCRF. \label{table:3}}  
\begin{center}   
\begin{tabular}{c|c|c}    
\hline    \textbf{\diagbox[width=15em]{Parameter}{Value}{Image Size}} & \textbf{1024$\times$1024} & \textbf{2048$\times$2048} \\
\hline $\omega^{(1)}$ & 40 & 40 \\
\hline $\omega^{(2)}$ & 3 & 3 \\
\hline $\theta^2_\alpha$ & 3 & 8 \\
\hline $\theta^2_\beta$ & 8 & 12 \\
\hline $\theta^2_\gamma$ & 3 & 3 \\
\hline iterations & 10 & 10 \\
\hline
\end{tabular}
\end{center}  
\end{table}

\subsection{Comparative Experiment}
\label{sec:3.2}

We quantify the semantic segmentation performance using the intersection over union (IoU, \citealt{Rezatofighi2019}) and F1-score \citep{David2011}, both of which are derived from the confusion matrices. All the metrics of MORDEN are calculated on the test set corresponding to its training dataset. We conducted three independent trainings with MHAS and AHAS and calculated their error bars. To facilitate comparison, we collected the corresponding metrics of the proposed method on each dataset, and trained our model once. As shown in Table~\ref{table:4}, our model outperforms these methods on their respective datasets. We made significant progress compared to U-Net \citep{Zheng2024} and Flat U-Net \citep{Zhu2025}, though not as much compared to Attention U\textsuperscript{2}-Net \citep{Jiang2024}. It is important to note that we trained our model with only one-sixth of the dataset used for Attention U\textsuperscript{2}-Net. Therefore, the results should be considered a relative reference rather than a controlled architectural comparison. Nevertheless, the overall increase demonstrates the effectiveness of our model. 
 
\renewcommand{\arraystretch}{1.0}
\begin{table}[h!]   
\caption{Metrics of semantic segmentation. \label{table:4}} 
\begin{center}   
\begin{tabular}{c|c|c|c} 
\hline    \textbf{Dataset} & \textbf{Model} & \textbf{IoU} & \textbf{F1-Score} \\
\hline    \multirow{2}{*}{HAS} & U-Net & 0.670 & 0.809 \\
\cline{2-4}     & MORDEN & \textbf{0.732} & \textbf{0.844} \\
\hline    \multirow{2}{*}{\citeauthor{Zhu2025}} & Flat U-Net & 0.68 & 0.81 \\
\cline{2-4}     & MORDEN & \textbf{0.789} & \textbf{0.879} \\
\hline    \multirow{2}{*}{\citeauthor{Jiang2024}} & Attention U\textsuperscript{2}-Net & 0.714 & 0.832 \\
\cline{2-4}     & MORDEN & \textbf{0.723} & \textbf{0.839} \\
\hline    MHAS & MORDEN & 0.696$\pm$0.005 & 0.819$\pm$0.004 \\
\hline    AHAS & MORDEN & 0.812$\pm$0.007 & 0.895$\pm$0.004 \\
\hline   
\end{tabular} 
\end{center}   
\end{table}

\subsection{Ablation Study}
\label{sec:3.3}

To verify the effectiveness of all individual components we proposed, a series of ablation studies were conducted. Table~\ref{table:5} summarizes the impact of each innovation. Most of the innovations provide an impressive progress in metrics, offering an insight into their effectiveness. However, after three times of training, the influence of multihead appears relatively weak. We have two explanations for this: First, due to the limited computational resources, MORDEN only contains two heads, which may not be sufficient for a significant improvement in metrics. Second, although it is not obvious in the metrics under current conditions, we also report an enhancement in training efficiency. The average number of epochs required to achieve best metrics with multihead is 469, whereas that without multihead is 668. In addition, we removed all APP in MORDEN and add PPM or ASPP after the downsampling stage to further verify the effectiveness of the hierarchical deployment of APP. The results demonstrate that our hierarchical design outperforms those which simply apply multiscale information fusion on high-level features.

\begin{table}[ht!]   
\caption{Ablation results on MHAS dataset. \label{table:5}}  
\begin{center} 
\begin{threeparttable}
\begin{tabular}{c|c|c}
\hline \textbf{Models} & \textbf{IoU} & \textbf{F1-Score} \\
\hline MORDEN & 0.696$\pm$0.005 & 0.819$\pm$0.004 \\
\hline w/o IN\tnote{*} & 0.678 & 0.805 \\
\hline w/o large kernel & 0.663 & 0.795 \\
\hline w/o APP & 0.678 & 0.806 \\
\hline w/o PAG & 0.688 & 0.813 \\
\hline w/o branch block & 0.688 & 0.812 \\
\hline w/o multihead & 0.695$\pm$0.003 & 0.818$\pm$0.002 \\
\hline PPM & 0.683 & 0.809 \\
\hline ASPP & 0.678 & 0.805 \\
\hline
\end{tabular}
 
\begin{tablenotes}
    \footnotesize
    \item[*] w/o denotes without.
\end{tablenotes}
\end{threeparttable}
\end{center}   
\end{table}

\subsection{Validation on Dataset Expansion}
\label{sec:3.4}

There are two major concerns regarding the expanded AHAS dataset. One is that does DenseCRF really optimize the edge details, and the other is that can the expansion of dataset truly imply more semantic information and facilitate deep learning models.

\subsubsection{Quantitative Assessment of Edge Details}
\label{sec:3.4.1}

We used the gradient magnitude and orientation to evaluate the effectiveness of DenseCRF. Considering each extracted fragment as a simply connected domain $D$ whose boundary is represented as $\partial{D}$, with the unit tangent vector $\vec{t}$ of every pixel belonging to $\partial{D}$, the average edge gradient magnitude and orientation are formulated as follows:

\begin{equation}
    \mu_{mag} = \frac{1}{\lvert \partial{D} \rvert}
    \sum_{(i, j) \in \partial{D}} \lVert \nabla I(i, j) \rVert_n,
\end{equation}
\begin{equation}
    \mu_{ori} = \frac{1}{\lvert \partial{D} \rvert}
    \sum_{(i, j) \in \partial{D}} \mathrm{arccos} \Bigl(
    \frac{\lvert \nabla I(i, j) \cdot \vec{t}(i, j) \rvert }{\lVert \nabla I(i, j) \rVert} \Bigr),
\end{equation}
where $I$ denotes the intensity of the original image, and $n$ specifies the distance norm, either L1 norm or L2 norm. For the convenience of comparison, the angle range is standardized to [0\degree, 90\degree]. These two values offer an intuitive measurement of the contrast on either side of the fragment boundaries: the higher they are, the finer the edge details should be. We calculated the metrics of the ground truth, the semantic segmentation masks and the post-processing results under image sizes of 1024$\times$1024 and 2048$\times$2048. We uniformly adopted the intensity of original 2K resolution images for consistency. As for extracting $D$, we applied the ``findcontours'' function in OpenCV which comes from \citet{SUZUKI1985}. The unit tangent vector $\vec{t}$ is obtained through calculating the slope of each pixel and its eight neighboring pixels belonging to $D$. Although there exist fitting errors, this method is effective for statistical comparison because it is a standardized processing procedure.

As shown in Table~\ref{table:6}, DenseCRF performs superbly at refining edge details. First, the improvement from the segmentation masks to the post-processed results under 1K resolution indicates the effectiveness of DenseCRF. Second, the 2K resolution results obtained by DenseCRF surpass the manually labeled ground truth, underscoring its solid efficacy. Third, the increase in metrics from 1K to 2K resolution suggests an information gap between different pixel resolutions and highlights the necessity of DenseCRF for creating a well-annotated dataset. Finally, the large-scale evaluation on the AHAS training set enhances the credibility of the insights.

\begin{table}[ht!]   
\caption{Average edge gradient magnitude and orientation of multiple masks.\label{table:6}}  
\begin{center} 
\begin{tabular}{c|c|c|c|c}
\hline \textbf{Dataset} & \textbf{Mask} & $\mu_{mag}$(L1) & $\mu_{mag}$(L2) & $\mu_{ori}$ \\
\hline \multirow{4}{*}{\shortstack{MHAS\\[5pt]test set}} & ground truth & 250.6 & 198.0 & 69.6 \\
\cline{2-5} & segmentation & 217.7 & 171.6 & 60.9 \\
\cline{2-5} & DenseCRF(1K) & 222.2 & 175.3 & 62.7 \\
\cline{2-5} & DenseCRF(2K) & \textbf{259.0} & \textbf{204.5} & \textbf{72.9} \\
\hline \multirow{3}{*}{\shortstack{AHAS\\[5pt]training set}} & segmentation & 179.1 & 140.7 & 58.8 \\
\cline{2-5} & DenseCRF(1K) & 180.9 & 142.1 & 59.6 \\
\cline{2-5} & DenseCRF(2K) & \textbf{208.1} & \textbf{163.8} & \textbf{66.8} \\
\hline
\end{tabular} 
\end{center}   
\end{table}

\subsubsection{Analysis of Data Scaling}
\label{sec:3.4.2}

We re-trained MORDEN with two specially designed dataset settings to estimate the effect of dataset expansion. Both settings use the validation and the test set of MHAS but differ in their training sets. One uses the AHAS training set, and the other uses the semantic segmentation masks without post-processing of the same 775 images in AHAS.

Compared to MORDEN trained on the MHAS training set, both of the two large-scale dataset settings demonstrate relative advancement, as shown in Table~\ref{table:7}. The raise in metrics under either settings is satisfactory and is expected to result from the expansion of the dataset and the inclusion of more  semantic information, such as difficult samples. However, it is noteworthy that the metrics under AHAS with DenseCRF are lower than those under AHAS without DenseCRF. As discussed above in Section~\ref{sec:3.4.1}, DenseCRF results are more reliable, and the distribution of AHAS should differ from MHAS, cautioning us against the illusion of model bias expansion. More specifically, models trained on the small-scale MHAS dataset may overfit some biased patterns, and expanding the data scale also magnifies such bias, resulting in high metrics on the test set with similar bias. Therefore, we tend to regard the metrics under AHAS with DenseCRF as a more conservative estimate, or as more aligned with the actual benefits of data scaling. For rigor, the true influence of dataset expansion should fall within the interval between the two results, with one serving as the lower bound, and the other as the upper bound.

\begin{table}[ht!]   
\caption{Metrics under different dataset settings. \label{table:7}}
\begin{center} 
\begin{tabular}{c|c|c|c}
\hline \textbf{Training set} & \textbf{Validation and test set} & \textbf{IoU} & \textbf{F1-Score} \\
\hline MHAS & MHAS & 0.696$\pm$0.005 & 0.819$\pm$0.004 \\
\hline AHAS & MHAS & 0.699$\pm$0.002 & 0.820$\pm$0.001 \\
\hline AHAS(w/o DenseCRF) & MHAS & 0.702$\pm$0.002 & 0.822$\pm$0.002 \\
\hline
\end{tabular} 
\end{center}   
\end{table}

\subsection{Expert Judgment for Fragment Integration}
\label{sec:3.5}

Integrating filament fragments on an H$\alpha$ image can be challenging and may lead to conflicting judgments among different individuals \citep{Hao2018}. Rather than evaluating the performance of DBSCAN using a ground truth, we invited three relevant researchers to evaluate the clustering results on the MHAS test set independently and then averaged their scores. We only counted a cluster that perfectly corresponded to a single filament as a hit. Cluster consisting of more than one filament or two clusters belonging to the same filament were both considered as failed clustering. Of all the 510 clusters, the number of correct clusters judged by the researchers are 409, 449, and 434 respectively, for an average ratio of 0.844. This is a relatively satisfactory result given our stringent criteria for correctness.

\subsection{Workflow Visualization}
\label{sec:3.6}

\begin{figure}[ht!]
    \centering
    \includegraphics[width=0.6\linewidth]{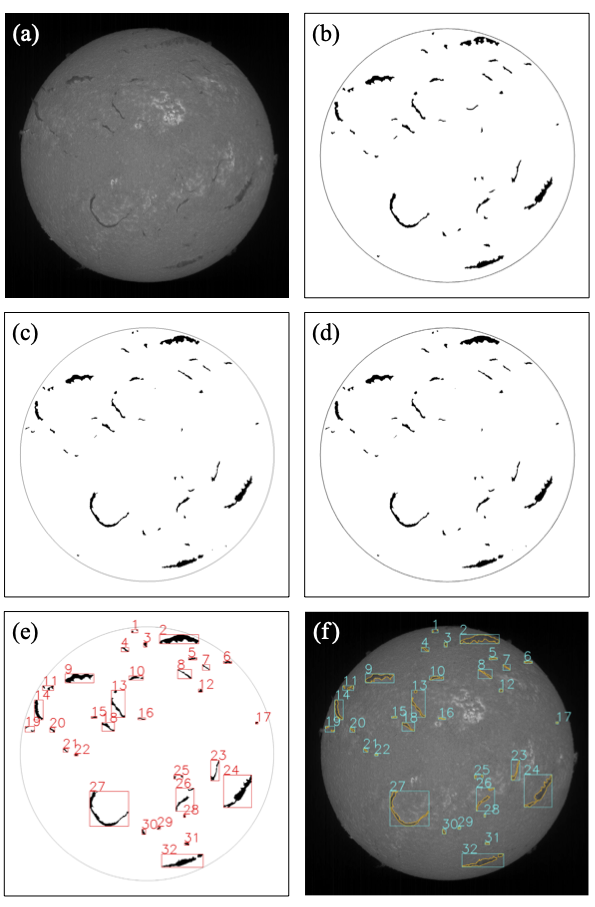}
    \caption{Visualization of the results of our workflow. (a) Original image. (b) Ground truth. (c) Semantic segmentation mask. (d) Post-processed mask under 2K resolution. (e) Cluster result. (f) Contour detail. The cycles in (b) -- (e) represent the solar limb for a clearer demonstration.}
    \label{fig:2}
\end{figure}

Figure~\ref{fig:2} shows an example of the workflow for detecting solar filaments. Figure~\ref{fig:2}(a) shows the original observed image selected from the MHAS test set.  The corresponding ground truth is shown in  Figure~\ref{fig:2}(b). Figure~\ref{fig:2}(c) shows the results detected directly by our MORDEN model. The post-processed mask shows the results derived from 2K resolution, as shown in  Figure~\ref{fig:2}(d). The semantic segmentation results demonstrate the high completeness of filaments, and the post-processed mask reflects the efficacy of DenseCRF in refining the filament boundaries. Notably, the polar crown filament near the north pole,  marked as No. 2 in Figure~\ref{fig:2}(e), is effectively refined when comparing to the detected result in  Figure~\ref{fig:2}(c) and the ground truth in Figure~\ref{fig:2}(b). Figure~\ref{fig:2}(e) shows the clustering results, where fragments of filaments belonging to a single filament are clustered together and marked by a light blue boundary box with a unique ID number. To provide a clear and intuitive view, the final results in Figure~\ref{fig:2}(e) overlap the original image. Filaments and filament fragments are marked with yellow contours.

\begin{figure}[htb!]
    \centering
    \includegraphics[width=0.6\linewidth]{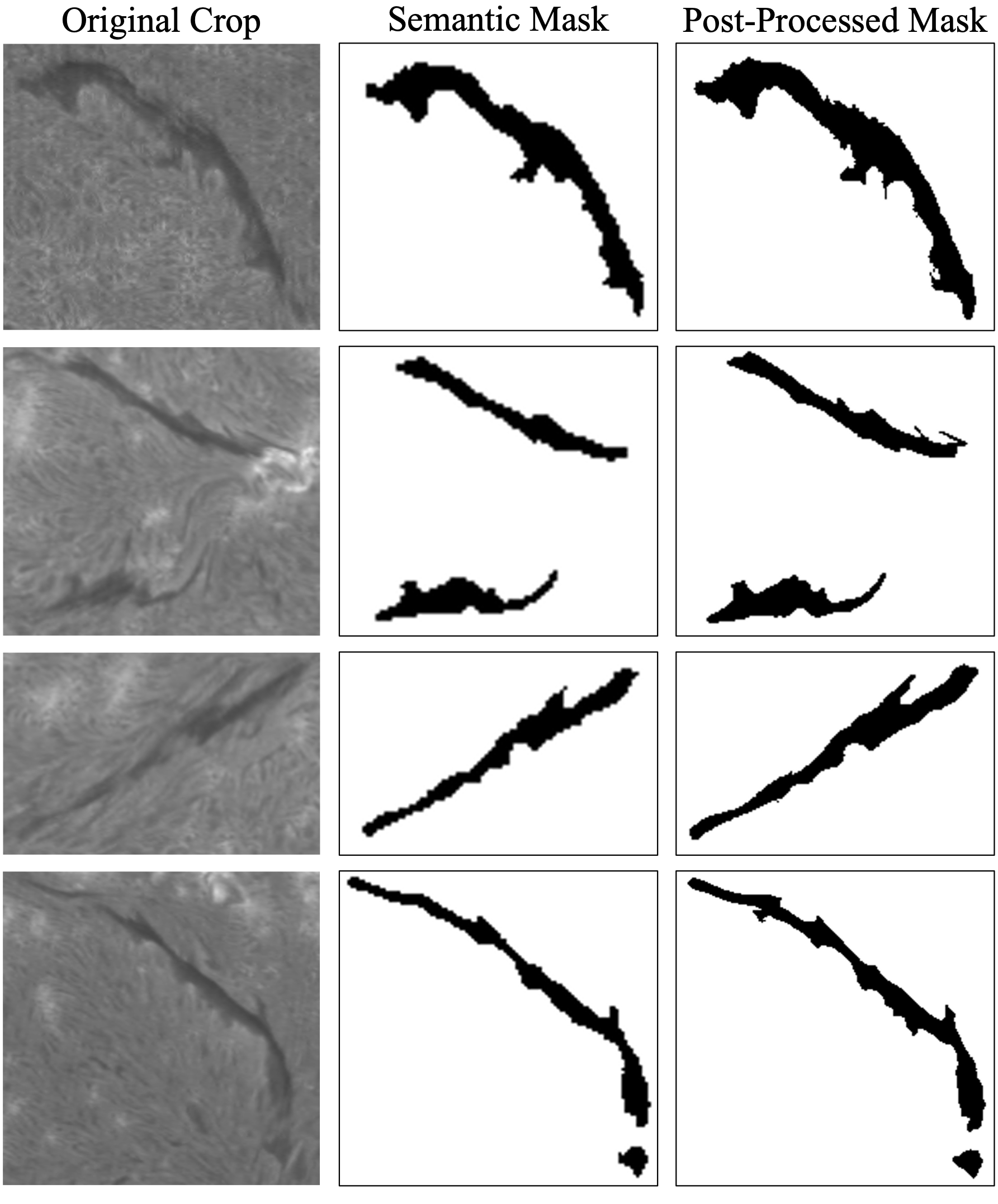}
    \caption{Enlarged detail of multiple examples for showing performance of post-processing stage.}
    \label{fig:3}
\end{figure}

In order to show the details of the results after post-processing, Figure~\ref{fig:3} gives several filament examples cropped from the full disk images. Compared to the semantic mask generated directly by MORDEN, the filament masks processed by DensCRF are smoother and the barb structures are more distinguishable.

\subsection{Visualization Examples for Robustness}
\label{sec:3.7}

\begin{figure*}[htb!]
    \centering
    \includegraphics[width=1.0\linewidth]{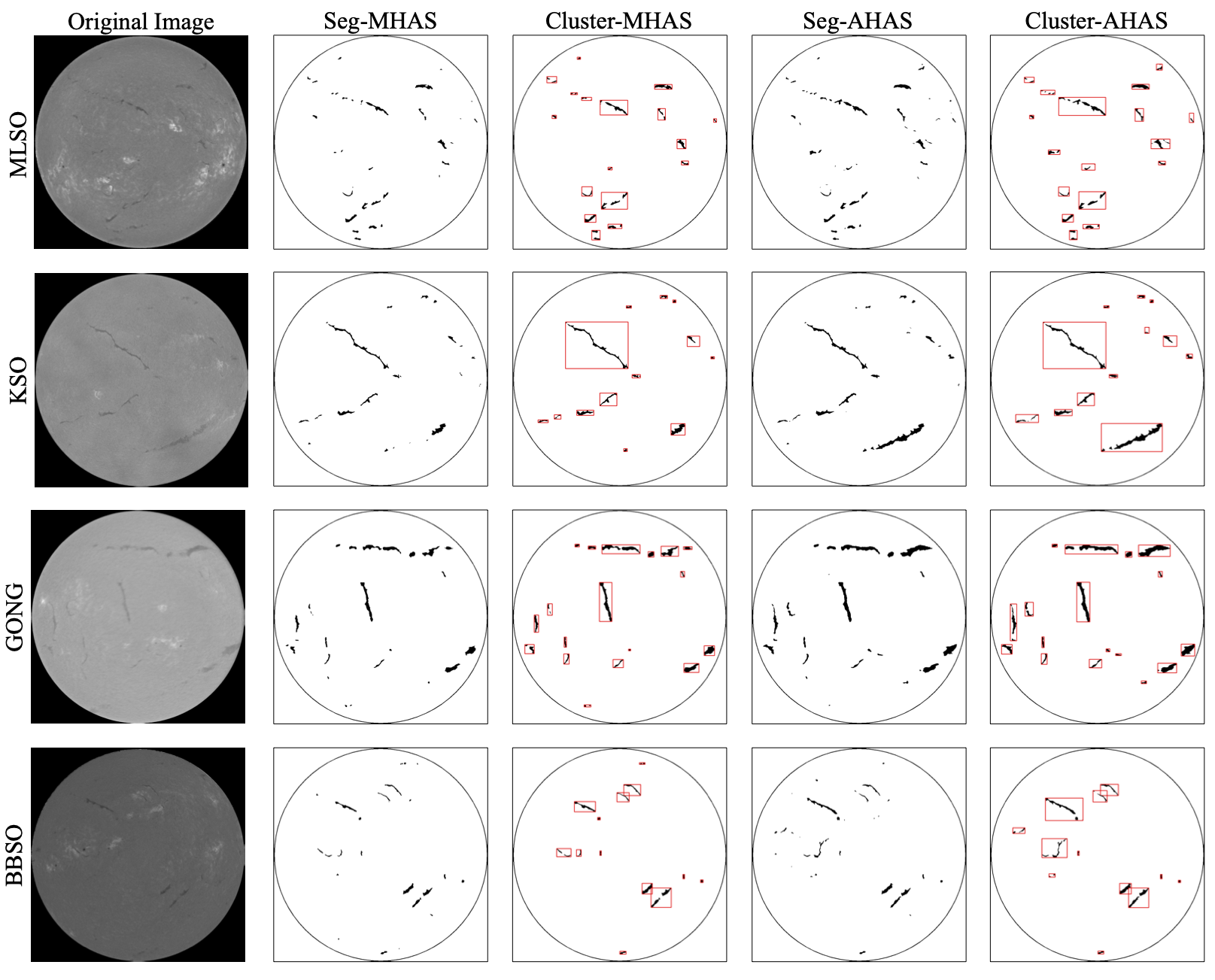}
    \caption{Transfer performance. The original images from top to bottom is observed at 17:05:57 UT on 17 September 2002 by MLSO, 07:42:45 UT on 26 May 2015 by KSO, 10:02:02 UT on 25 June 2025 by GONG, and 18:13:54 UT on 13 March 2013 by BBSO, respectively.}
    \label{fig:4}
\end{figure*}

To estimate the robustness of our methods, we collected observations from multiple different observatories and ran our workflow on them without any additional training. The images all possess several typical observational defects, varying in brightness, contrast, signal-to-noise ratio, and blurring. Figure~\ref{fig:4} shows that our methods adapt to different selected image qualities without encountering any obstacles. For example, the image selected from the Mauna Loa Solar Observatory (MLSO) shows a relatively low pixel resolution; the image selected from the Kanzelh\"ohe Observatory (KSO) is distorted by atmospheric clouds; the image selected from Global Oscillation Network Group (GONG) has a relatively low resolution; and the image selected from Big Bear Solar Observatory (BBSO) has a relatively low contrast. The segmentation model succeeded to extract most of the filaments with few misjudgment, and a majority of the clustering results are reasonable. This suggests a relatively high robustness in complex observational environments. 

Additionally, MORDEN trained on AHAS performs significantly better than MORDEN trained on MHAS, which is encouraging. For example, MORDEN trained on AHAS detects the filament in the central region of the northern hemisphere of the MLSO image, the quiescent filament in the southwest of the KSO image, the filament running north-south near the eastern hemisphere of the GONG image, and the relatively large filament with a north-east direction in the northern hemisphere of the BBSO image, as shown in Figure~\ref{fig:4}. These results favorably compare with the corresponding filament in the MHAS data and further corroborate previous quantitative findings on data scaling. More visualization results are available at \url{https://github.com/irisaltHu/MORDEN}.

\section{Discussion of Limitations}
\label{sec:4}
\subsection{Failure of Segmentation}
\label{sec:4.1}

Although our manually annotated dataset and multiscale feature extraction model ensure complete filament detection, the semantic segmentation model still fails to recognize some small filaments in active regions or near the edge of solar disk, where the projection effects are significant. We have identified several underlying causes of this issue. First, these two types of filaments are inherently semantically ambiguous, which makes it easy to introduce individual bias during annotation. For example, as instruments continue to advance and resolution improves, increasingly fine filament structures would become visible. However, we have also observed more fibrils, which makes it difficult to determine whether they are filaments, especially in active regions. Some edge filaments are difficult to label due to their small inclination angle (i.e., nearly parallel to the latitude lines), which causes their true dimensions to appear very small due to projection effect. This may also be because parts of the structures lie near the solar limb, such as prominences. Second, despite continuous architectural innovations \citep{ConvNext2022}, the current entirely CNN-based architecture remains outperformed by prominent state-of-the-art architectures in computer vision where Transformer is nearly indispensable. This also reveals the third issue that the essential requirement for training transformer-based models, i.e., a large-scale, high-quality dataset, remains unattainable in this domain. As demonstrated in our dataset construction methodology, creating such datasets through manual annotation is both resource-intensive and inherently imprecise, resulting in limitations in the ground truth. Therefore, while our work does not represent a definitive solution to these complex problem of automated solar filament detection, it establishes a methodological foundation for advancing deep learning approaches through rigorous data quality assessment and architecture innovation.

\subsection{Applicability of AHAS}
\label{sec:4.2}

The occasional failure of the segmentation model raises another question about the applicability of AHAS. Our multiple experiments have demonstrated the validity of the training set of AHAS, and the drawbacks of segmentation are unlikely to substantially impact the training set, as a considerable portion of hard samples will still be included. However, the validation and test sets must be able to provide unbiased assessments of model performance, which is a much higher requirement. We are concerned that the construction of the AHAS validation and test set is not capable of revealing the efficacy of models under actual observational environments. Consequently, when employing AHAS, we recommend incorporating more credible data into the validation and the test sets. For example, manually annotating several images with hard samples and refining the details with DenseCRF would reinforce credibility.

To further verify the impact of this limitation, we conducted a simple experiment with our trained models. First, we applied DBSCAN to the MHAS test set to identify all filaments. Second, we counted the number of filaments with at least 10\% of their area recognized. The results in Table ~\ref{table:8} reveal that MORDEN, when trained on and chosen by MHAS, recognized 430 filaments out of 475 in the MHAS test set. When trained on the AHAS training set and chosen by the MHAS validation set, MORDEN recognized 435 filaments. However, when trained on and chosen by AHAS, it recognized only 421 filaments. This indicates that the missed detections in the AHAS training set will not severely weaken the performance of the model as long as the validation set is solid, possibly due to the large sample size. However, it is also clear that the AHAS validation set is not ideal for selecting a model during training.

\renewcommand{\arraystretch}{1.0}
\begin{table}[ht!]
\caption{Experiment on the applicability of AHAS \label{table:8}}  
\begin{center}   
\begin{tabular}{c|c|c|c|c}    
\hline    \textbf{Training set} & \textbf{Validation set} & \textbf{Filament number} & \textbf{Hit number} & \textbf{Hit rate} \\
\hline MHAS & MHAS & 475 & 430 & 0.905 \\
\hline AHAS & MHAS & 475 & 435 & 0.916 \\
    \hline AHAS & AHAS & 475 & 421 & 0.887 \\
\hline
\end{tabular}
\end{center}  
\end{table}

\subsection{Drawbacks of DenseCRF}
\label{sec:4.3}

Despite its excellent performance, DenseCRF has two notable drawbacks. First, its efficiency is relatively lower than that of CNNs. Rough statistical calculations show that MORDEN takes 0.17 seconds to process one image at 1K resolution, whereas MORDEN with DenseCRF takes 2 (6.5) seconds to process one image at 1K (2K) resolution, making it not preferred for real-time deployment. Second, DenseCRF requires solid intensity information, indicating a low robustness to poor image quality and noise. Therefore, DenseCRF may not be applicable for low quality historical images. Thus, the realistic employment of DenseCRF will encounter more foreseeable difficulties than laboratory validation, which is precisely why we use it for dataset generation. Due to the superior efficiency and robustness of deep learning models, AHAS dataset inherits the compelling efficacy of DenseCRF while avoiding its shortcomings.

\subsection{Suboptimal Fragment Integration}
\label{sec:4.4}

The integrity of filament fragments remains not fully addressed in this work. Due to the projection effects, determining whether the fragments belong to a single filament using the the H$\alpha$ image alone is difficult because of the three-dimensional configuration of the filament. This may require incorporating additional data, such as multi-wavelength or stereoscopic observations. However, acquiring such observations from historical data is limited, which hinders the feasibility for long-term statistical analysis. Accordingly, DBSCAN was deployed as a suboptimal compromise to strike a balance between usability and effectiveness.

\section{Conclusion}
\label{sec:5}

We presented a solar filament detection workflow designed for superior performance and robustness. Our workflow focuses on two primary challenges: the multiscale feature extraction task with long-tail distribution and the absence of large-scale, highly complete, and finely detailed dataset. To address the first challenge, we proposed the semantic segmentation model MORDEN, which is primarily reinforced for multiscale receptive field inspired by several modern innovations in computer vision. Next, we employed and validated DenseCRF to refine the edge details of segmentation results and pave the way for constructing a large-scale, high-quality dataset. As a result, we successfully built a fully automated annotated dataset AHAS. Finally, we applied DBSCAN for fragment integration to complete the solar filament detection workflow.

Through detailed experiments and analyses of each stage and critical issue in the workflow, we estimated the efficacy and limitations of each process and their production, deduced the cause of the imperfections, and suggested the solutions. The contribution of our methods is summarized as follows. First, MORDEN is an effective semantic segmentation model for solar filament within current datasets. It ensures the complete extraction of solar filaments and can be used for actual observations in various observational environments. Second, the post-processed results of DenseCRF are satisfactory for finer filament boundaries because it utilizes the intensity features of high-resolution original images to the maximum extent. However, MORDEN with DenseCRF is not the ultimate solution for solar filament segmentation since MORDEN is not ideal enough for perfect completeness in filament recognition, and the construction of AHAS may pave the way for the use of Transformer-based deep learning models. Furthermore, we expect that deep learning models with AHAS will overcome the inefficiencies of DenseCRF. Finally, we reiterate that using DBSCAN for fragment integration is suboptimal, primarily due to the semantic ambiguity of solitary H$\alpha$ image.

The advantages and the shortcomings of our methods provides pathways for our future work. With our current workflow, it is feasible to conduct a long-term statistical analysis of solar filaments with high robustness and credibility. In this case, depending on historical H$\alpha$ image is reasonable. On the other hand, we can integrate stereoscopic observations to achieve the best possible solar filament detection results with the current observational capabilities.

\section*{Data Accessibility}
\label{sec:6}

The corresponding access to the collected observations are provided below.

\textbf{MLSO}: Data downloaded from \url{https://www2.hao.ucar.edu/mlso}.

\textbf{KSO}: Data downloaded from \url{http://cesar.kso.ac.at}.

\textbf{GONG}: Data downloaded from \url{https://gong.nso.edu}.

\textbf{BBSO}: Data downloaded from \url{http://www.bbso.njit.edu}.

\textbf{CHASE}: Data downloaded from \url{https://ssdc.nju.edu.cn/NdchaseSatellite}.

The codes of the whole workflow and more visualization results are available at \url{https://github.com/irisaltHu/MORDEN}, and the datasets can be accessed through \url{https://zenodo.org/records/21582245}.

\begin{acknowledgements}
We are grateful to the CHASE, BBSO, KSO, GONG, and MLSO teams for providing the observational data. The present work was supported by NSFC under grants 12173019, 12333009, 12127901, the CNSA project D050101, the Fundamental Research Funds for the Central Universities 14380065, KG202506, and the Young Data Scientist Program of the China National Astronomical Data Center, as well as the AI \& AI for Science Project of Nanjing University.
\end{acknowledgements}


\bibliography{manuscript}{}
\bibliographystyle{raa}
\label{lastpage}

\end{document}